\documentclass[10pt]{iopart}
\usepackage{graphicx}% Include figure files
\usepackage{cite}
\begin{document}

\title[Ising model in generalized triangular lattices]{Monte Carlo study of frustrated Ising model with nearest- and next-nearest-neighbor interactions in generalized triangular lattices}

\author{Hoseung Jang and Unjong Yu$^{*}$}

\address{Department of Physics and Photon Science \&
         Research Center for Photon Science Technology,
          Gwangju Institute of Science and Technology, 
          Gwangju 61005, South Korea}
\eads{\mailto{uyu@gist.ac.kr} (corresponding author)}

\date{\today}
%\preprint{AIP/123-QED}

\begin{abstract}
We investigate the frustrated $J_1$-$J_2$ Ising model with nearest-neighbor interaction $J_1$ and next-nearest-neighbor interaction $J_2$ in two kinds of generalized triangular lattices (GTLs) employing the Wang--Landau Monte Carlo method and finite-size scaling analysis. In the first GTL (GTL1), featuring anisotropic properties, we identify three kinds of super-antiferromagnetic ground states with stripe structures. Meanwhile, in the second GTL (GTL2), which is non-regular in next-nearest-neighbor interaction, the ferrimagnetic 3$\times$3 and two kinds of partial spin liquid ground states are observed. We confirm that residual entropy is proportional to the number of spins in the partial spin liquid ground states. Additionally, we construct finite-temperature phase diagrams for ferromagnetic nearest-neighbor and antiferromagnetic next-nearest-neighbor interactions. In GTL1, the transition into the ferromagnetic phase is continuous, contrasting with the first-order transition into the stripe phase. In GTL2, the critical temperature into the ferromagnetic ground state decreases as antiferromagnetic next-nearest-neighbor interaction intensifies until it meets the 3$\times$3 phase boundary. For intermediate values of the next-nearest-neighbor interaction, two successive transitions emerge: one from the paramagnetic phase to the ferromagnetic phase, followed by the other transition from the ferromagnetic phase to the 3$\times$3 phase.
\end{abstract}


%%%%%%%%%%%%%%%%%%%%%%%%%%%%%%
% WARNING:
% (1) The words table and figure should be written in full and not abbreviaged to tab. and fig. Do not include ‘eq.’, ‘equation’ etc before an equation number or ‘ref.’ ‘reference’ etc before a reference number.
% (2) Note it is not normally necessary to include the word equation before an equation number except where the number starts a sentence.
% (3) Roman d for a differential d, a Roman e for an exponential e and a Roman i for the square root of −1. (Use \rmd, \rme, and \rmi.)

\section{Introduction}

Geometric frustration arises when competing interactions between individual elements, such as spins or particles, cannot be simultaneously satisfied due to the lattice's geometric constraints~\cite{Greedan01,Moessner06,Lacroix11}. It has attracted significant attention owing to a variety of ground states~\cite{Starykh15,Yu15PRE,Yu16,Lhotel20}, unusual phase transitions\cite{Lhotel20,Vojta18,Yahne21}, and potential importance in diverse research fields including unconventional superconductivity~\cite{Ardavan12,Chen13}, spintronics~\cite{Lohani19,Haley23}, quantum computing~\cite{Lopez-Bezanilla23,Zhao24}, and battery technologies~\cite{Duvel17,Irvine22}.

In the context of the Ising model~\cite{Ising25}, which is a cornerstone in statistical mechanics, geometric frustration occurs when there exist loops composed of odd-numbered antiferromagnetic interactions. In two-dimensional antiferromagnetic Ising lattice systems, among the eleven Archimedean and twenty 2-uniform lattices, the kagome and triangular lattices have the strongest frustration effects, facilitating the realization of spin liquid ground state, where the spins fluctuate dynamically even at absolute zero temperature without settling into any particular ordered arrangement~\cite{Yu15PRE,Yu16}. In less frustrated lattices, diverse ground states are observed: partial spin ice, spin ice, partial long-range-ordered (partial spin liquid), and long-range-ordered states~\cite{Yu15PRE,Yu16}.

The inclusion of next-nearest-neighbor interactions ($J_2$) in addition to nearest-neighbor interactions ($J_1$) of the conventional Ising model further enriches phase diagrams. In the $J_1$-$J_2$ Ising model, geometric frustration arises when at least one of the nearest- and next-nearest-neighbor interactions is antiferromagnetic. In the case of the square lattice---typically free from frustration in the conventional Ising model---a super-antiferromagnetic ground state (collinear order) appears when the antiferromagnetic next-nearest-neighbor interaction surpasses half of the nearest-neighbor interaction in magnitude ($|R|>1/2$ with $R=J_2/J_1$)~\cite{Landau80,Ramazanov16}. It is generally accepted that the phase transition into the ferromagnetic or antiferromagnetic phase is continuous and belongs to the two-dimensional Ising universality class~\cite{Ramazanov16,Li21,Yoshiyama23}. In the phase transition into the collinear super-antiferromagnetic phase, on the other hand, tricriticality has been proposed: the phase transition is first-order for $1/2<|R|<R^{*}$ while it is continuous with the weak Ashkin--Teller universality beyond the tricritical point ($|R|>R^{*}$)~\cite{Jin12,Jin13,Yoshiyama23}. (However, there is still some controversy about the location and presence of the tricritical point. See, e.g., reference~\cite{Gangat24}.) Conversely, the Ising antiferromagnet in the triangular lattice, which has spin liquid ground state by severe frustration, exhibits the $\sqrt{3}$$\times$$\sqrt{3}$ ferrimagnetic and collinear super-antiferromagnetic ground states for ferromagnetic and antiferromagnetic next-nearest-neighbor interactions, respectively~\cite{Mihura77,Brandt86}. It is notable that the phase transition from the high-temperature paramagnetic phase into the low-temperature collinear super-antiferromagnetic phase was reported to be first-order~\cite{Rastelli05,Malakis07}. Intriguingly, reentrant phenomena and successive phase transitions have been observed in the $J_1$-$J_2$ Ising models in generalized kagome~\cite{Azaria87}, generalized cubic~\cite{Yokota89}, generalized honeycomb~\cite{Diep91}, and generalized square~\cite{Debauche92} lattices, where some portion of next-nearest interactions are eliminated. Although conjectured to be related to partially disordered phases~\cite{Diep91,Debauche92}, the origin and mechanism of such unusual phase transitions remain unclear.

In this paper, we propose two kinds of generalized triangular lattices and study their properties using the Wang--Landau method with finite-size scaling analysis. From the entropy profiles, ground state phase diagrams and residual entropies are calculated. We identify various ground states: ferromagnetic, collinear, partial order (partial spin liquid), and 3$\times$3 ferrimagnetic phases. Furthermore, we construct finite-temperature phase diagrams for ferromagnetic nearest-neighbor interaction and antiferromagnetic next-nearest-neighbor interaction. We report first-order and successive phase transitions in addition to continuous phase transitions. The outline of this paper is as follows. In section~\ref{Sec_Model}, the model Hamiltonian of the $J_1$-$J_2$ Ising model~\cite{Ising25}, two kinds of generalized triangular lattices, and the calculational methods (the Wang--Landau method and finite-size scaling) are introduced. Section~\ref{Sec_Results} presents the results for the two lattices and discusses their implications. Finally, section~\ref{Sec_Summary} is for a summary of the results and concluding remarks.

\begin{figure*}[tb]
%\centering
\includegraphics[width=0.5\columnwidth]{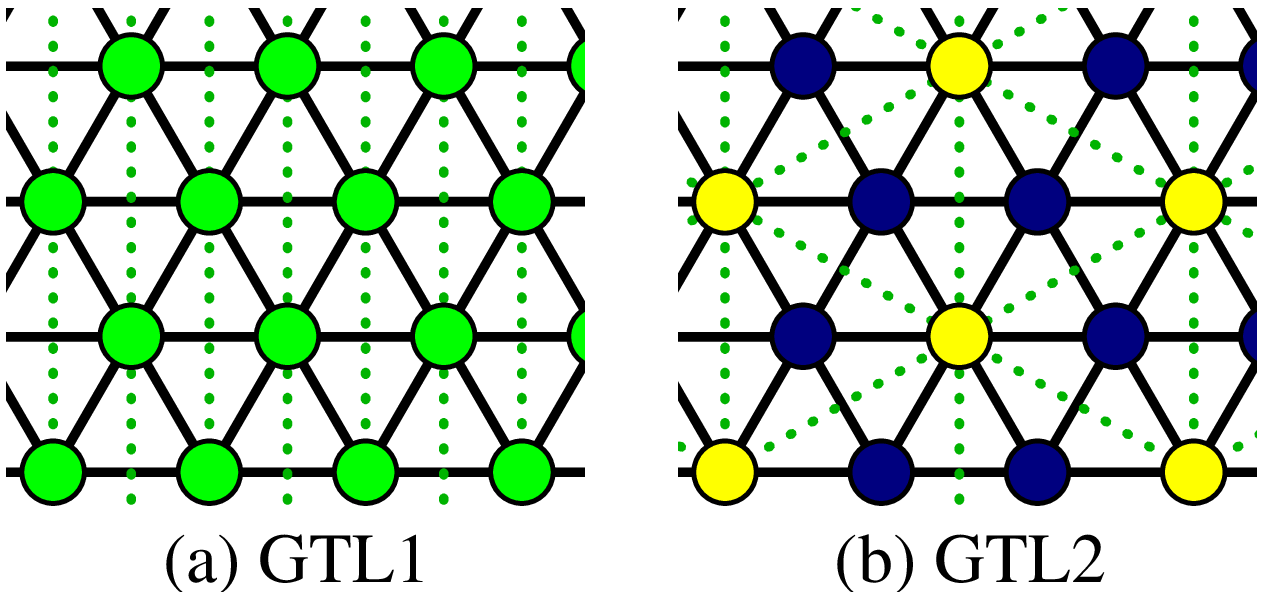}
\caption{Two generalized triangular lattices (GTLs) examined in this work. Circles indicate the locations of the Ising spins. Black solid lines and green dotted lines represent interactions between nearest-neighbor spins ($J_1$) and between next-nearest-neighbor spins ($J_2$), respectively. In (b), yellow and navy blue circles denote the six-next-nearest-neighbor spins and no-next-nearest-neighbor spins, respectively. It is noteworthy that GTL1 is spatially anisotropic and regular while GTL2 is isotropic and non-regular.}
\label{fig1}
\end{figure*}

\section{Model and methods \label{Sec_Model}}

The $J_1$-$J_2$ Ising model~\cite{Ising25,Mihura77} under investigation is described by the Hamiltonian
\begin{eqnarray}
H = -J_1 \sum_{\langle i,j \rangle} s_i s_j - J_2 \sum_{\langle\!\langle i,j \rangle\!\rangle} s_i s_j , \label{Hamiltonian_Ising}
\end{eqnarray}
where the first and the second summations run over all nearest- and next-nearest-neighbor spin pairs, respectively, excluding double counting. The Ising spin $s_i$ at site $i$ has either an up or down spin direction, denoted by $s_i=\pm 1$. Positive (\textit{resp.} negative) values of $J_1$ and $J_2$ represent ferromagnetic (\textit{resp.} antiferromagnetic) interactions. Geometric frustration occurs if and only if at least one of $J_1$ and $J_2$ is negative.

In this work, we consider the two-dimensional generalized triangular lattice (GTL), where the nearest-neighbor interaction is the same as the conventional triangular lattice while some of the next-nearest-neighbor interactions are missing. The first GTL studied in this work (GTL1) is spatially anisotropic, with the next-nearest-neighbor interactions present only in one direction while $J_2=0$ in the other two directions. Consequently, every spin interacts with six nearest-neighbors and two next-nearest-neighbors. It is worth mentioning that another anisotropic triangular lattice, which has next-nearest-neighbor interaction only in two directions, had been proposed~\cite{Kitatani88}. Since it has the $\sqrt{3}$$\times$$\sqrt{3}$ ground state for $J_1<0$ and $J_2>0$ just like the isotropic triangular lattice, it had been examined in relation to the Kosterlitz--Thouless transition~\cite{Kitatani88,Miyashita91,Queiroz95,Otsuka06}. In the second GTL (GTL2), two types of spins exist; the next-nearest-neighbor interactions exist only between the first type of spins, which occupy one of tripartite sublattices. Hence, GTL2 is non-regular; 1/3 spins have six next-nearest-neighbors while the other 2/3 spins have no next-nearest-neighbors. Refer to figure~\ref{fig1}, which describes the two kinds of lattices. In this study, we investigate the $J_1$-$J_2$ Ising model on these two GTLs to explore the impact of anisotropy and non-regularity on the phase diagram and phase transitions in the frustrated $J_1$-$J_2$ Ising model. Each lattice comprises $N=L^2$ spins, where $L$ denotes the linear size of the lattice. The periodic boundary condition is adopted in all directions.

Given the highly frustrated nature of this model, we employ the Wang--Landau algorithm~\cite{Wang01}, which is free from the critical and supercritical slowing-down~\cite{Janke94} and yields reliable results even in the presence of strong frustration. It has been proven to be efficient to determine entropy profile, which restricts equilibrium phases and non-equilibrium dynamics, in frustrated systems~\cite{Silva06,Ramazanov16,Azhari20,Azhari22,Jin22,Azhari23}.

The Wang--Landau algorithm calculates the density of states $\rho(E_1, E_2)$ through a random walk in energy space $(E_1, E_2)$, where $(-J_1E_1)=-J_1(\sum_{\langle i,j \rangle} s_i s_j)/N$ and $(-J_2E_2)=-J_2(\sum_{\langle\!\langle i,j \rangle\!\rangle} s_i s_j)/N$ are the energies per spin by the interactions between nearest and next-nearest spin pairs, respectively. The transition probability from configuration $I$ with $\left(E_{1}^{(I)}, E_{2}^{(I)}\right)$ to configuration $F$ with $\left(E_{1}^{(F)}, E_{2}^{(F)}\right)$ is given by
\begin{eqnarray}
P(I\rightarrow F)
= \mbox{Min}\left[1, \frac{\rho\left(E_{1}^{(I)}, E_{2}^{(I)}\right)}{\rho\left(E_{1}^{(F)}, E_{2}^{(F)}\right)}\right].
\end{eqnarray}
At each step, the density of states $\rho(E_{1}, E_{2})$ is multiplied by $\lambda_n$, which is an empirical factor larger than 1, and the histogram $h(E_1,E_2)$ increases by one. As the simulation progresses, the histogram tends to be flat. Once the standard deviation of $h(E_1,E_2)$ is less than 4\% of average $h(E_1,E_2)$ for all accessible ordered pair $(E_{1}, E_{2})$, a new set of random walks begins with an empty histogram and an updated $\lambda_{n+1} = \sqrt{\lambda_{n}}$. The value of $\lambda_n$ begins from $\lambda_0 = e$ and the simulation terminates if $\lambda_n < \exp(10^{-10})$. As the final outcome, $\rho(E_1, E_2)$ is determined. After the normalization of $\sum_{E_1, E_2}\rho(E_1, E_2) = 2^N$, the entropy of the states $(E_1, E_2)$ can be obtained by $S(E_1,E_2)=k_B\log[\rho(E_1,E_2)]$~\cite{Yu15PRE,Yu16,Jin22}. We set the Boltzmann constant $k_B=1$ for convenience throughout this paper. The residual entropy (zero-point entropy), which is the entropy at zero temperature, is $S_0 = S(E_1^{(g)},E_2^{(g)})$, where  $E_1^{(g)}$ and $E_2^{(g)}$ are the values of $E_1$ and $E_2$ in the ground state. The size of the energy space is proportional to $z_1 z_2 L^4$, where $z_1$ and $z_2$ are the average numbers of nearest- and next-nearest-neighbors of each spin, respectively. Consequently, the maximum attainable lattice size in this work is limited to $L=36$ in GTL1 and $L=42$ in GTL2.

The ensemble average of any thermodynamic variable $\langle O \rangle(T,J_1,J_2)$ can be calculated for given $\rho(E_1, E_2)$ and $O(E_{1}, E_{2})$ for any values of $J_1$ and $J_2$ by 
\begin{eqnarray}
\langle O \rangle (T,J_1,J_2) = \frac{\sum_{E_1, E_2} O(E_{1}, E_{2}) \rho(E_1, E_2) \, \rme^{(J_1 E_1 + J_2 E_2)/T}}{\sum_{E_1, E_2} \rho(E_1, E_2) \, \rme^{(J_1 E_1 + J_2 E_2)/T}}.
\end{eqnarray}
The average value $O(E_{1}, E_{2})$ for the states with $(E_{1}, E_{2})$ can be obtained by additional Wang--Landau iterations after $\rho(E_1, E_2)$ is determined. The whole calculation was repeated five times for each $L$, and the results were averaged. Computational details are elaborated in \cite{Silva06,Azhari20}.

Using the Wang--Landau method, we calculated the ensemble average of magnetization $m$, susceptibility $\chi$, Binder cumulant $U$~\cite{Binder81}, and specific heat $C$ as a function of temperature $T$. They are defined as
\begin{eqnarray}
m = \frac{1}{N} \left\langle \left| \sum_{i=1}^{N} s_i \right| \right\rangle, \\
\chi = \frac{N}{T} \left( \left\langle m^2 \right\rangle - \left\langle m \right\rangle^2 \right), \label{eq_sus} \\
U = 1 - \frac{\left\langle m^4 \right\rangle}{3\left\langle m^2 \right\rangle^2}, \label{eq_U}\\
C = \frac{N}{T^2} \left( \left\langle E^2 \right\rangle - \left\langle E \right\rangle^2 \right), 
\end{eqnarray}
where $E=\langle H \rangle / N$ denotes energy per spin and $\langle \cdots \rangle$ means the ensemble average. The magnetization ($m$) acts as the order parameter for the ferromagnetic ordering. For a ground state with different ordering, a suitable order parameter must be defined, and the susceptibility and Binder cumulant should be calculated using the suitable order parameter in place of $m$.

To determine the critical temperature $T_c$, we employed two methods based on finite-size scaling~\cite{Binder81,Binder81PRL}. In the former method, the location of the crossing point of Binder cumulant data of different sizes provides the critical temperature~\cite{Binder81PRL,Challa86,Azhari20,Azhari22}. In the latter method, $T_c$ is obtained by the fitting to 
\begin{eqnarray}
\left[ T_c^{(L)} - T_c \right] \propto L^{-1/\nu} , \label{nu}
\end{eqnarray}
where $\nu$ is the critical exponent of the correlation length. The pseudo critical temperature $T_c^{(L)}$ is the temperature that maximizes susceptibility or specific heat in the lattice of size $L$. Throughout this work, we employed the standard least-squares fitting with the chi-squared test. Note that these two methods are applicable not only to continuous phase transitions but also to first-order phase transitions~\cite{Binder81PRL,Challa86}. We mainly used the former method, which is more precise in general~\cite{Yu15}, and the latter method was used only for comparison purposes. The correction-to-scaling~\cite{Ferrenberg91,Yu15}, which gives only minor corrections in the finite-size scaling in this model, is ignored.

\section{Results and Discussion \label{Sec_Results}}

\begin{figure*}[tb]
%\centering
\includegraphics[width=0.7\columnwidth]{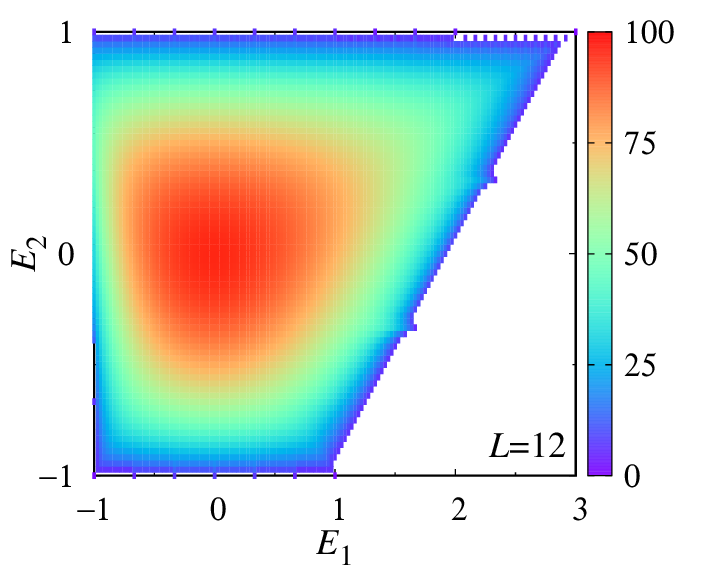}
\caption{Entropy profile as a function of $E_1=(\sum_{\langle i,j \rangle} s_i s_j)/N$ and $E_2=(\sum_{\langle\!\langle i,j \rangle\!\rangle} s_i s_j)/N$ for the $J_1$-$J_2$ Ising model in GTL1 of $L=12$.}
\label{fig2}
\end{figure*}

\begin{figure*}[tb]
%\centering
\includegraphics[width=0.7\columnwidth]{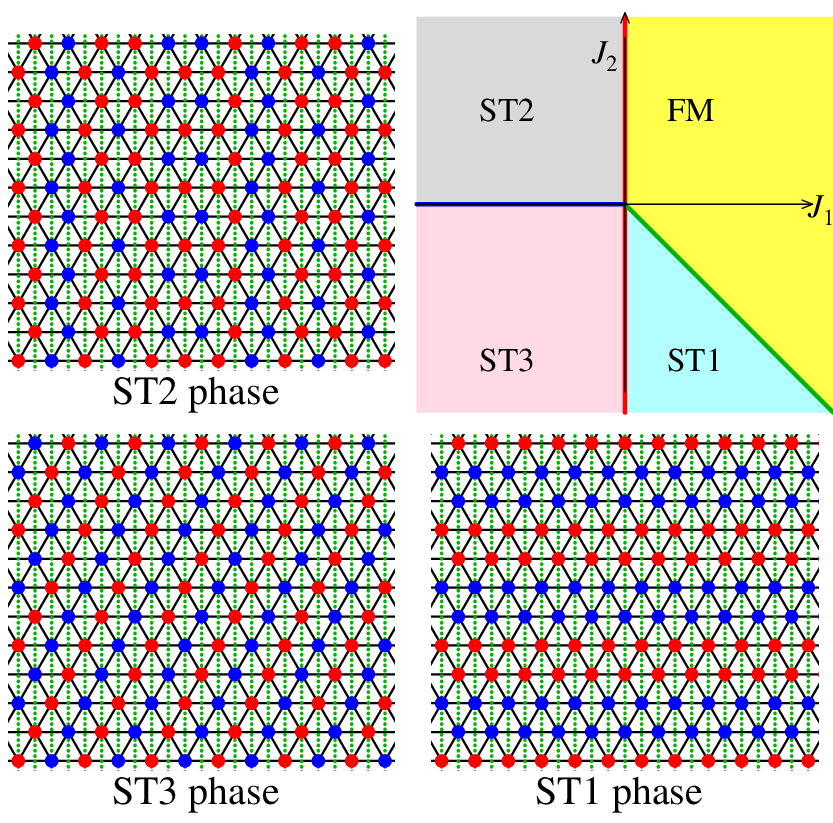}
\caption{Zero-temperature phase diagram as a function of $J_1$ and $J_2$ for the $J_1$-$J_2$ Ising model in GTL1. Yellow, light cyan, gray, and pink colors represent regions where the ground state is ferromagnetic (FM) and stripe (ST1, ST2, and ST3) phases, respectively. The green line indicates the phase boundary of $J_2 = -J_1$ with $J_1>0$. Typical spin configurations for each stripe ground state are presented, where blue and red circles represent up and down spins, respectively.}
\label{fig3}
\end{figure*}

\begin{figure}[bt]
%\centering
\includegraphics[width=0.5\columnwidth]{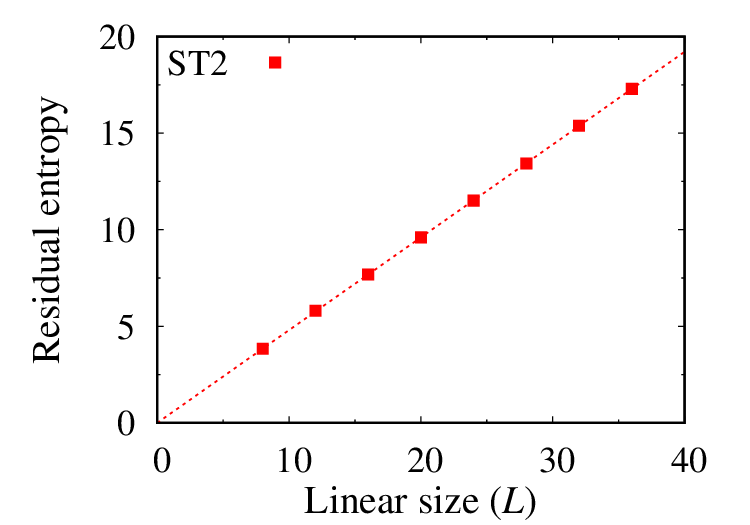}
\caption{Residual entropy ($S_0=\log[\rho(E_1=-1,E_2=1)]$) of the ST2 state, which is the ground state for $J_1 < 0$ and $J_2 > 0$ in GTL1. The statistical errors are smaller than the size of the symbols. The red straight line ($S_0 = (0.48027) L$) was obtained by fitting to $S_0 \propto L$.}
\label{fig4}
\end{figure}

\begin{figure*}[tb]
%\centering
\includegraphics[width=1.0\columnwidth]{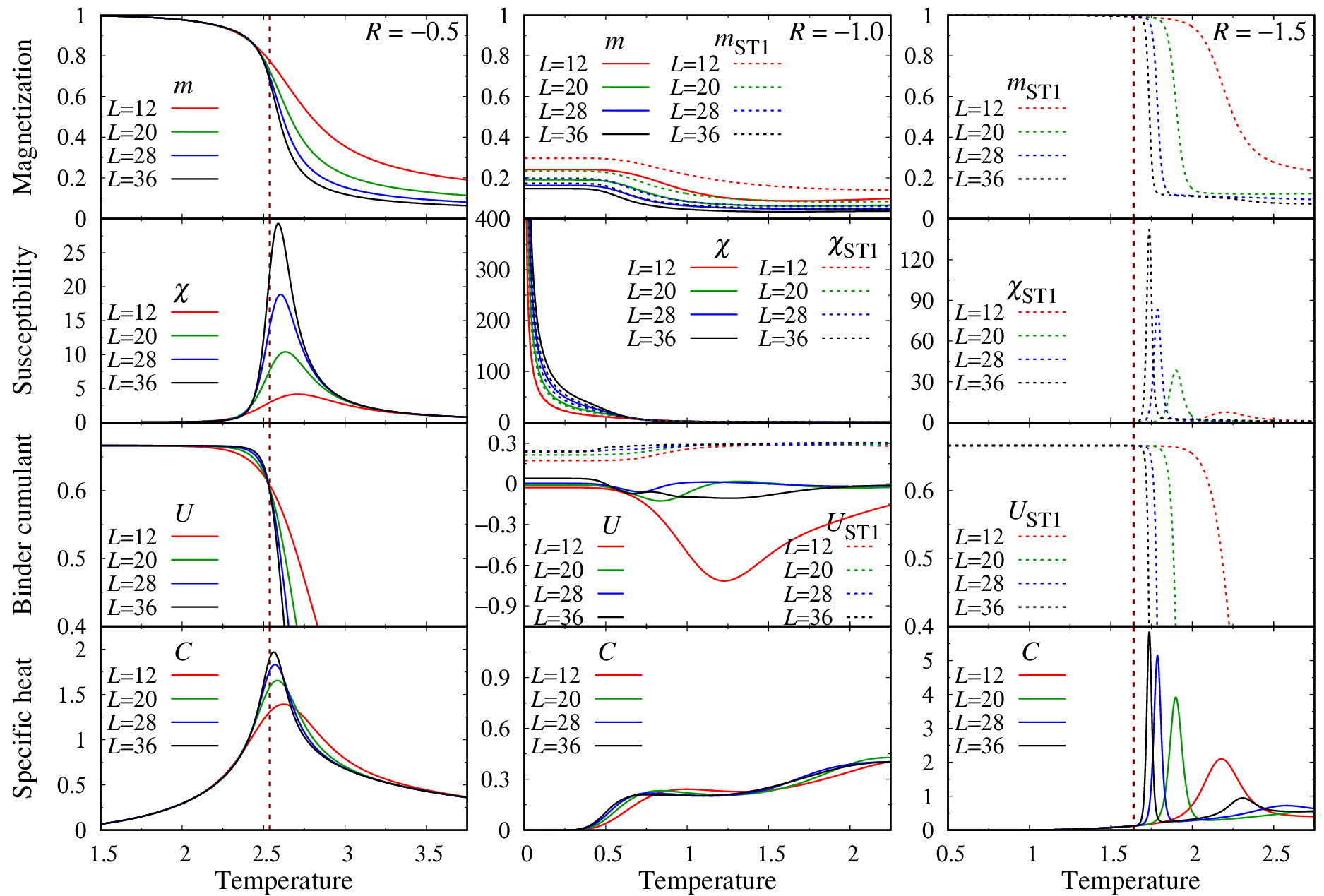}
\caption{Magnetization ($m$ and $m_{\mathrm{ST1}}$), magnetic susceptibility ($\chi$ and $\chi_{\mathrm{ST1}}$), Binder cumulant ($U$ and $U_{\mathrm{ST1}}$), and specific heat ($C$) as a function of temperature ($T$) for $J_1 = 1$ and $R = J_2 / J_1$ varying the lattice size $L$ in GTL1. For $R = \{-1.0, -1.5\}$, ST1-type ordering is assumed in $m_{\mathrm{ST1}}$, $\chi_{\mathrm{ST1}}$, and $U_{\mathrm{ST1}}$. The vertical dotted lines represent the values of the critical temperature of the infinite lattice obtained by the crossing of the Binder cumulant data for lattices of $L=32$ and $L=36$.}
\label{fig5}
\end{figure*}

\begin{figure}[tb]
%\centering
\includegraphics[width=1.0\columnwidth]{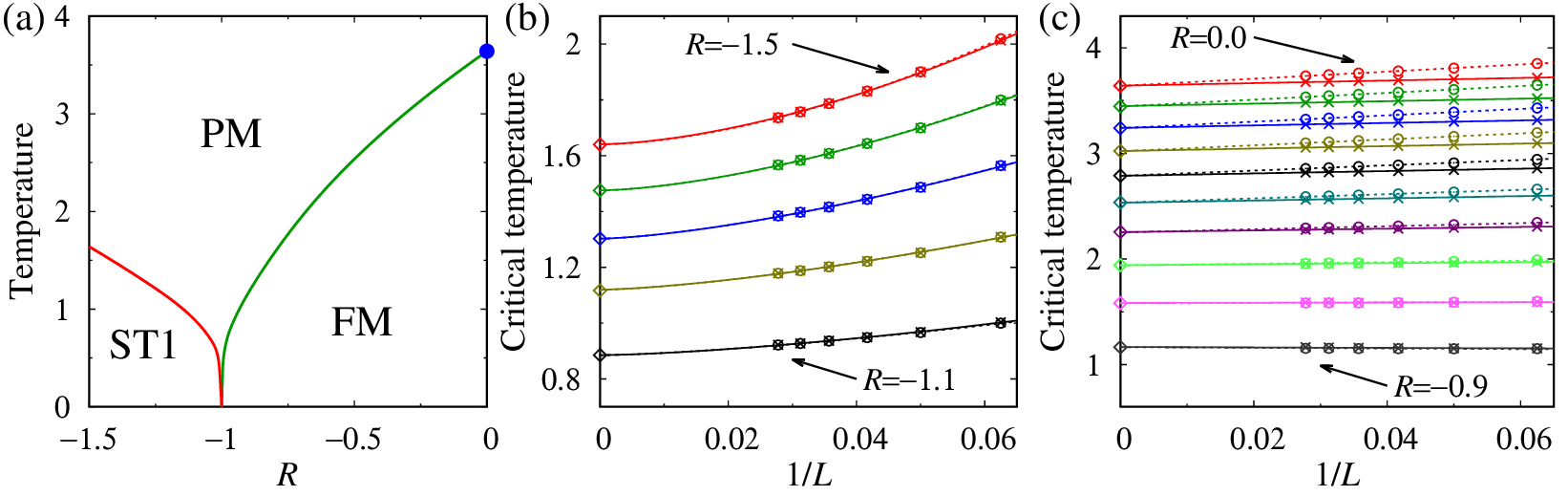}
\caption{(a) Finite temperature phase diagram for $J_1 = 1$ and $R = J_2 / J_1$ in GTL1. PM and FM denote paramagnetic and ferromagnetic phases, respectively. For the ST1 phase, refer to figure~\ref{fig3}. The exact value of $T_c=4/\log(3)$ at $R=0$ is indicated by the solid circle~\cite{Wannier50,Codello10}. The error bar is comparable to the line thickness. Pseudo critical temperature $T_c^{(L)}$ that is determined by maximum susceptibility ($\chi$ or $\chi_{\mathrm{ST1}}$) and specific heat ($C$) are plotted by empty circles ($\circ$) and crosses ($\times$), respectively, in (b) and (c). Lines in (b) and (c) are from fitting to $T_c^{(L)} = T_c + AL^{-1/\nu}$, where fitting parameters are $T_c$, $A$, and $\nu$. Diamond symbols ($\diamond$) at $1/L=0$ show the critical temperatures obtained by the Binder cumulant crossing. Values of $R$ are $-1.5$, $-1.4$, $-1.3$, $-1.2$, and $-1.1$ from top to bottom in (b); they are $0$, $-0.1$, $-0.2$, $-0.3$, $-0.4$, $-0.5$, $-0.6$, $-0.7$, $-0.8$, and $-0.9$ from top to bottom in (c).}
\label{fig6}
\end{figure}

\begin{figure*}[tb]
%\centering
\includegraphics[width=1.0\columnwidth]{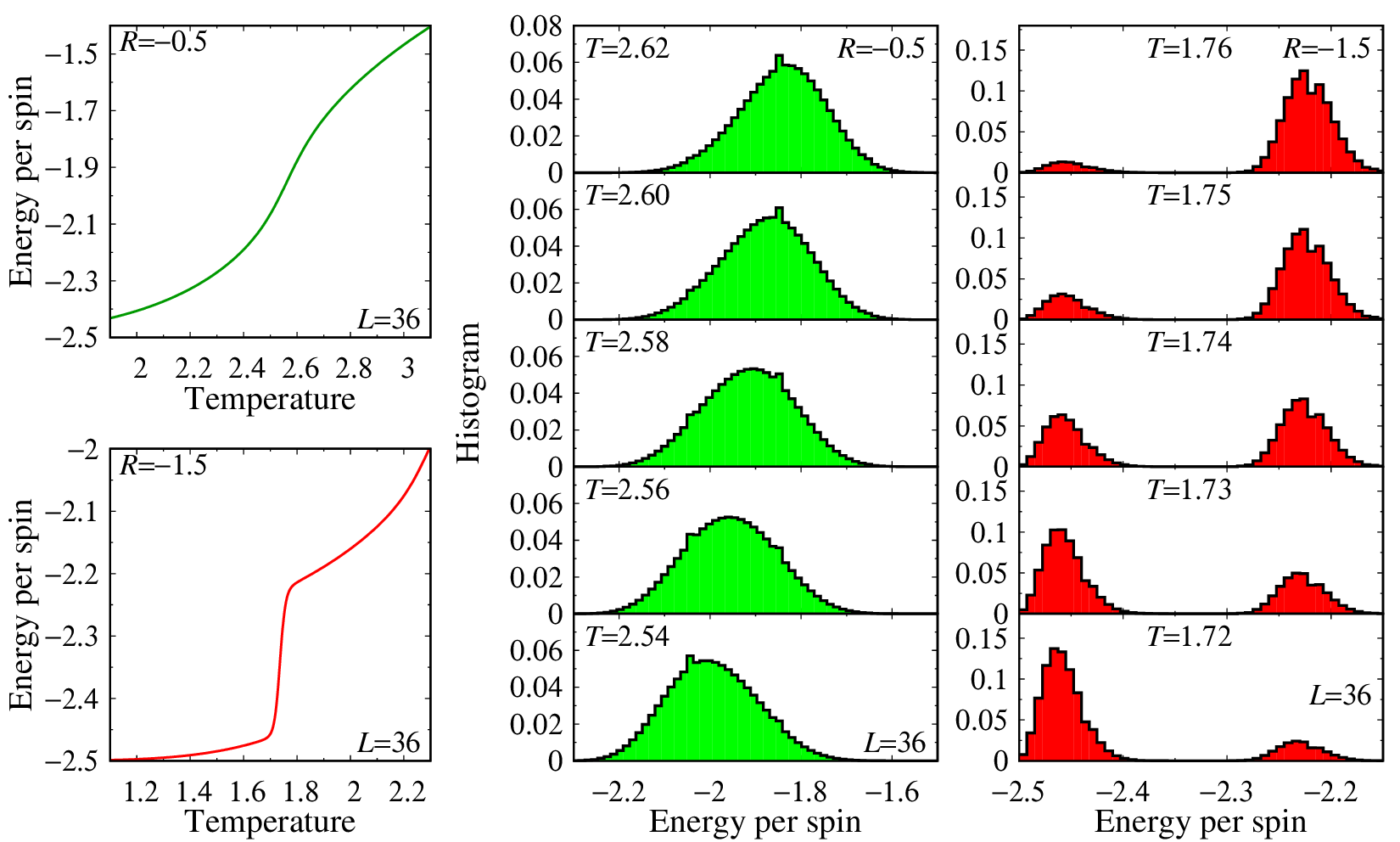}
\caption{The left panels depict energy per spin for $R=-0.5$ and $R=-1.5$, respectively, in GTL1. The middle and right panels show the histograms as functions of energy near the pseudo critical temperature $T_c^{(L)}$ with $L=36$ for $R=-0.5$ and $R=-1.5$.}
\label{fig7}
\end{figure*}

\subsection{$J_1$-$J_2$ Ising model in GTL1}

Figure~\ref{fig2} illustrates the entropy profile as a function of $E_1$ and $E_2$ for the $J_1$-$J_2$ Ising model in GTL1 obtained by the Wang--Landau method. The four corner points of the entropy profile $(E_1=3, E_2=1)$, $(E_1=1, E_2=-1)$, $(E_1=-1, E_2=1)$, and $(E_1=-1, E_2=-1)$ correspond to the ferromagnetic (FM) and three kinds of super-antiferromagnetic stripe ground states (ST1, ST2, and ST3), respectively. Given that the energy per spin is $E=-J_1 E_1 -J_2 E_2$, determining the ground state for given $J_1$ and $J_2$ is straightforward. For example, the FM phase is the ground state when its energy per spin ($E_{\mathrm{FM}}=-3J_1-J_2$) is lower than $E_{\mathrm{ST1}}=-J_1+J_2$, $E_{\mathrm{ST2}}=J_1-J_2$, and $E_{\mathrm{ST3}}=J_1+J_2$, implying $J_1>-J_2$ and $J_1>0$. The parameter regimes of $J_1$ and $J_2$ obtained by this method and typical ground state spin configurations for each stripe ground state are depicted in figure~\ref{fig3}.

For the FM ground state, a two-fold degeneracy arises from the spin up-down symmetry. The ST1 and ST3 ground states exhibit a four-fold degeneracy. In the case of ST2, the spins within each vertical stripe have the same spin direction; while two adjacent vertical stripes may have the same spin direction, three successive stripes of the same spin direction are prohibited. Consequently, long-range ordering exists only along the stripes, and the residual entropy is proportional to the number of stripes $L$ as shown in figure~\ref{fig4}. Notably, the proportional factor may vary depending on the lattice's shape and boundary conditions.

Our focus in this study lies on the fourth quadrant of the zero-temperature phase diagram in figure~\ref{fig2}, where the interactions between nearest-neighbor spins are ferromagnetic ($J_1 > 0$) and those between next-nearest-neighbor spins are antiferromagnetic ($J_2 < 0$). Without loss of generality, we fixed $J_1=1$ and calculated physical quantities as a function of $R=J_2/J_1$. Figure~\ref{fig5} shows magnetization ($m$ and $m_{\mathrm{ST1}}$), magnetic susceptibility ($\chi$ and $\chi_{\mathrm{ST1}}$), Binder cumulant ($U$ and $U_{\mathrm{ST1}}$), and specific heat ($C$) as a function of temperature ($T$) for $R = -0.5$, $-1.0$, and $-1.5$. The order parameter for the ST1 phase is defined by 
\begin{eqnarray}
m_{\mathrm{ST1}} = \frac{\mathrm{Max}\left\{ \left| M_0 + M_1 - M_2 - M_3 \right|, \left| M_0 - M_1 - M_2 + M_3 \right| \right\}}{N} ,
\end{eqnarray}
where $M_i$ represents the total magnetization of $(4n+i)$th rows with $n=\{0,1,\cdots,(L/4-1)\}$ ($M_i = \sum_{\{(x,y)|y \% 4 = i\}} s_{(x,y)}$, where $s_{(x,y)}$ is the spin in the $x$th column and $y$th row). The order parameter $m_{\mathrm{ST1}}$ becomes 1 at the ST1 ground state and $m_{\mathrm{ST1}}$ vanishes for the ferromagnetic or paramagnetic phases. Corresponding susceptibility ($\chi_{\mathrm{ST1}}$) and Binder cumulant ($U_{\mathrm{ST1}}$) are defined by replacing $m$ with $m_{\mathrm{ST1}}$ in (\ref{eq_sus})--(\ref{eq_U}).

At $J_2=0$ ($R=0$), an exact solution exists~\cite{Wannier50,Codello10}: a continuous phase transition from the paramagnetic phase into the ferromagnetic phase occurs at $T_c=4/\log(3)$. The left panel in figure~\ref{fig5} ($R=-0.5$) also exhibits typical features of a continuous phase transition. As shown in the finite-temperature phase diagram in figure~\ref{fig6}, the critical temperature decreases as $R$ decreases and finite-temperature phase transition disappears at $R=-1$; for $R<-1$, a different kind of phase transition is observed. The values of the critical temperature in figures~\ref{fig5}--\ref{fig6} were determined by the crossing of the Binder cumulant data for the two largest lattices studied in this work ($L=32$ and $L=36$). The middle and right panels of figure~\ref{fig6} confirm that the values of critical temperature are consistent with the extrapolation of the locations of maximum specific heat ($C$) and susceptibility ($\chi$ or $\chi_{\mathrm{ST1}}$) to $L\rightarrow \infty$ with (\ref{nu}) for $-1.5\leq R \leq 0$. 
The value of critical exponent $\nu$ is $1.7(1)$ for $R=-1.5$ and decreases monotonically to $\nu=1.4(1)$ as $R$ increases to $-1.1$. Although the uncertainty in $\nu$ is large, the transition temperature can be determined with good precision, as the value of $T_c$ obtained by fitting is not sensitive to the specific values of $\nu$.
For $-0.9\leq R \leq 0$, $\nu=1$ within statistical error bar. Note that $\nu=1$ in the two-dimensional Ising model without frustration~\cite{onsager1944}. Therefore, the results of figures~\ref{fig5}--\ref{fig6} suggest that the phase transitions for $R>-1$ are continuous phase transitions that belong to the two-dimensional Ising universality class while the nature of the phase transitions for $R<-1$ is different.

Figure~\ref{fig7} shows the energy per spin and its histogram near the pseudo critical temperature $T_c^{(L)}$ in the lattice of $L=36$ for $R=-0.5$ and $R=-1.5$. In the middle panel of figure~\ref{fig7} ($R=-0.5$), only one peak is evident, with the peak position continuously shifting to the lower energy as temperature decreases. This is a typical behavior of continuous phase transitions. To the contrary, a clear two-peak structure is observed for $R=-1.5$. As temperature decreases, the lower-energy peak rises while the higher-energy peak diminishes. This behavior verifies the first-order phase transition~\cite{Rastelli05,Cary17,Azhari22}.
This double-peak structure persists for $R<-1$, and the energy gap at the transition increases as $R$ decreases with no indication of tricriticality observed in the $J_1$-$J_2$ Ising model on the square lattice~\cite{Jin12,Jin13,Yoshiyama23}.

%%%%%%%%%%%%%%%%%%%%%%%%%%%%%%%%%%%%%%%%%%%%%%%

\begin{figure*}[tb]
%\centering
\includegraphics[width=0.7\columnwidth]{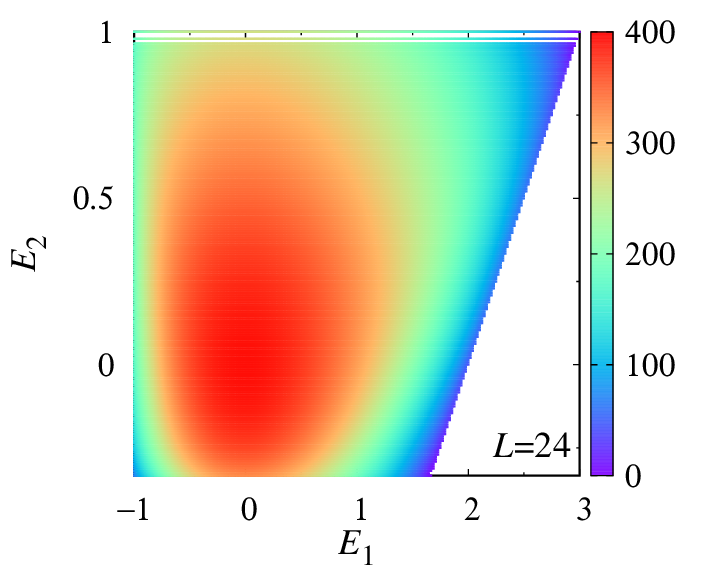}
\caption{Entropy profile as a function of $E_1$ and $E_2$ for the $J_1$-$J_2$ Ising model in GTL2 of $L=24$.}
\label{fig8}
\end{figure*}

\begin{figure*}[tb]
%\centering
\includegraphics[width=0.7\columnwidth]{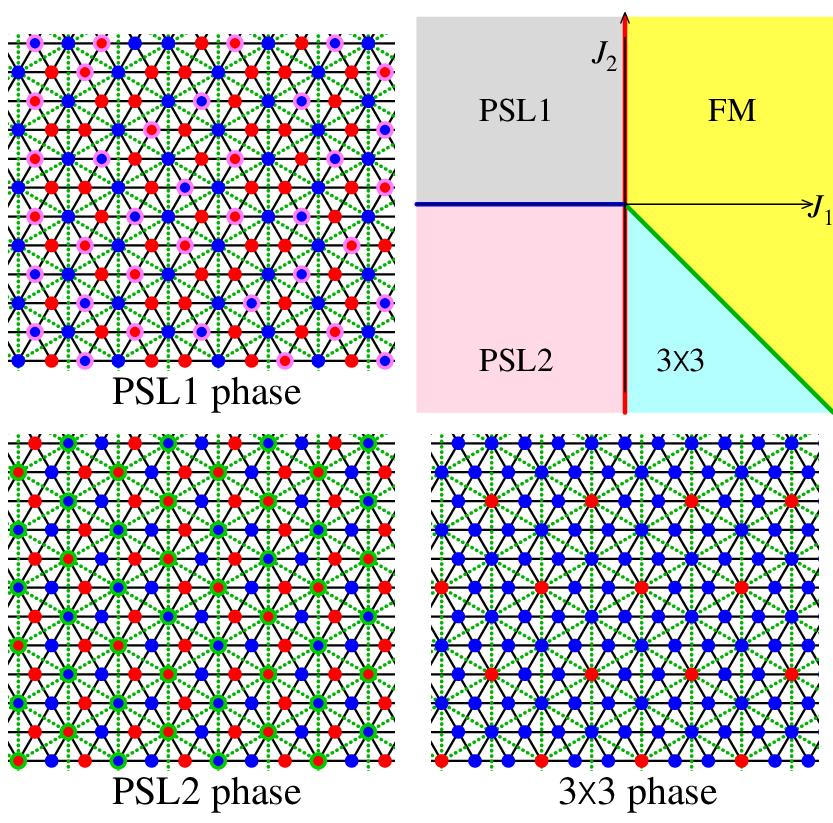}
\caption{Zero-temperature phase diagram for the $J_1$-$J_2$ Ising model in GTL2. Yellow, light cyan, gray, and pink colors represent regions where the ground states are the ferromagnetic (FM), 3$\times$3, and partial spin liquid (PSL1 and PSL2) phases, respectively. The green straight line indicates the phase boundaries of $J_2 = -J_1$ with $J_1>0$. Typical spin configurations for the 3$\times$3 and partial spin liquid ground states are illustrated; blue and red circles represent up and down spins, respectively. Violet circles in the PSL1 phase indicate spins that can overturn their directions without changing the total energy, while green circles in the PSL2 phase indicate six-next-nearest-neighbor spins.}
\label{fig9}
\end{figure*}

\begin{figure}[bt]
%\centering
\includegraphics[width=0.5\columnwidth]{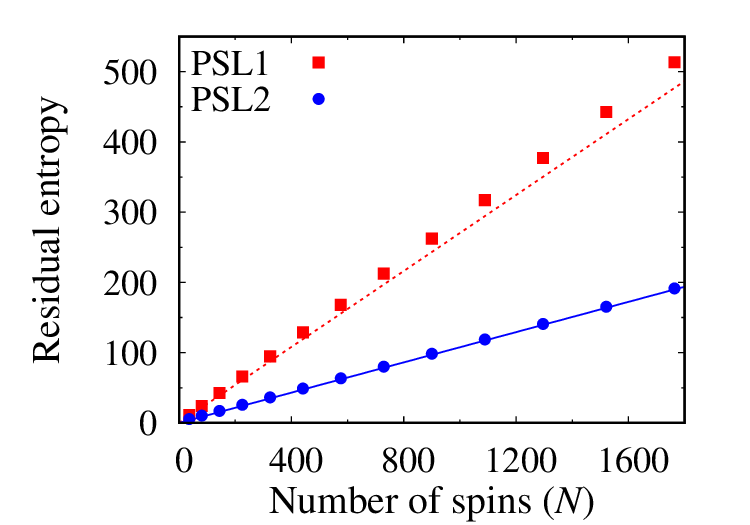}
\caption{Residual entropy ($S_0=\log[\rho(E_1=-1,E_2=1)]$ for the PSL1 state and $S_0=\log[\rho(E_1=-1,E_2=-1/3)]$ for the PSL2 state) of the two partial spin liquid states, which are the ground states for ($J_1 < 0$, $J_2 > 0$) and ($J_1 < 0$, $J_2 < 0$), respectively, in GTL2. The statistical errors are smaller than the size of the symbols. The red dotted line represents an estimate by the Pauling's approximation ($S_0 = (2/3)\log[3/2] N$), which is 7\% smaller than the numerical results. The blue line shows $S_0 = (\sigma_0^{T}/3) N$, where $\sigma_0^{T} \approx 0.323066$ is the residual entropy per spin of the antiferromagnetic Ising model in the triangular lattice~\cite{Wannier50}.
}
\label{fig10}
\end{figure}

\begin{figure*}[tb]
%\centering
\includegraphics[width=1.0\columnwidth]{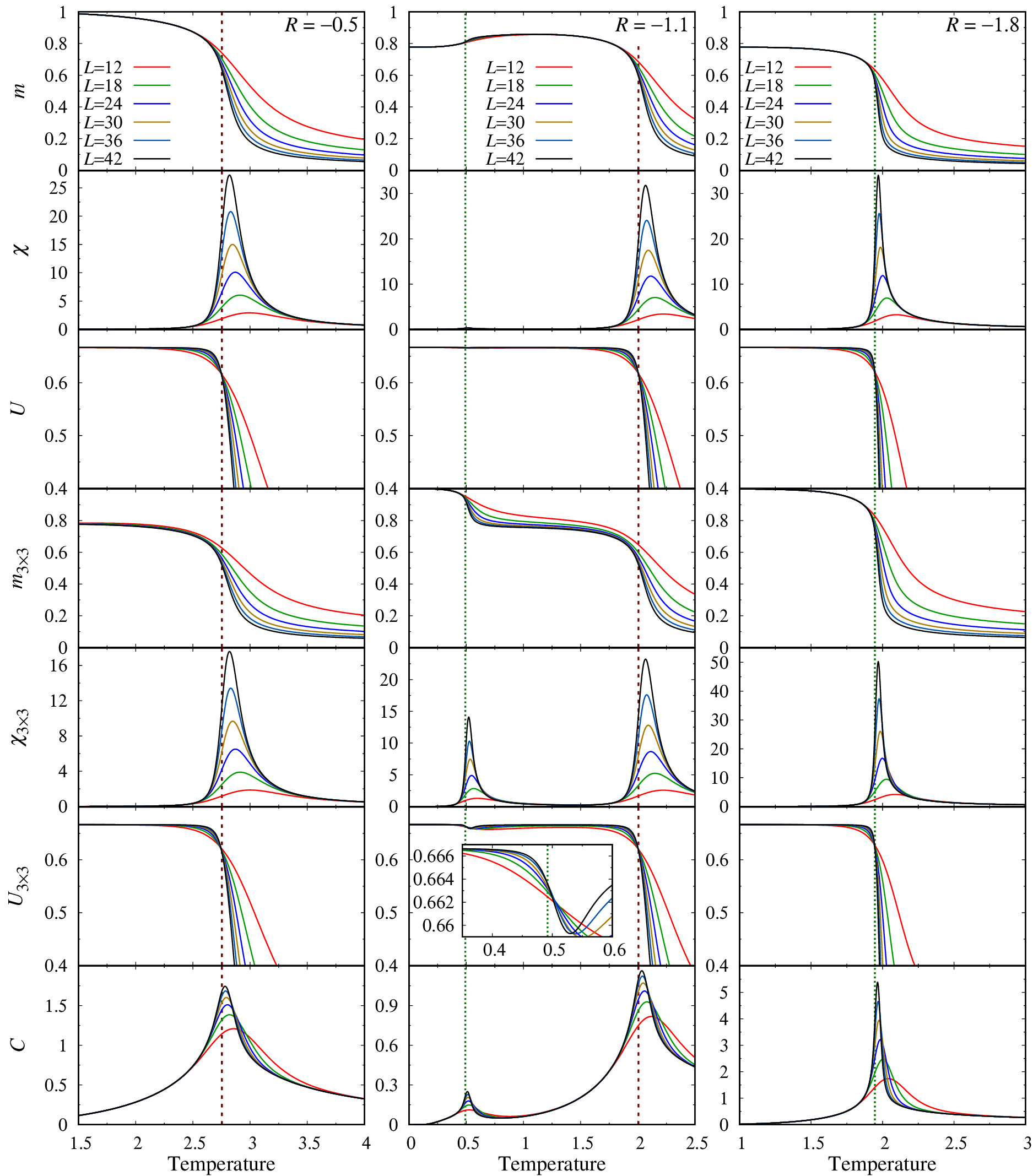}
\caption{Magnetization ($m$ and $m_{3\times 3}$), magnetic susceptibility ($\chi$ and $\chi_{3\times 3}$), Binder cumulant ($U$ and $U_{3\times 3}$), and specific heat ($C$) as a function of temperature ($T$) for $J_1 = 1$ and $R = J_2 / J_1 =\{-0.5,-1.1, -1.8\}$ varying the lattice size $L$ in GTL2. The vertical dotted lines indicate the values of the critical temperature of the infinite lattice obtained by the crossing of the Binder cumulant data ($U$ for transitions into the ferromagnetic phase and $U_{3\times 3}$ for transitions into the 3$\times$3 phase) for lattices of $L=36$ and $L=42$.  The inset in the middle panel magnifies $U_{3\times 3}$ near the lower-temperature phase transition.}
\label{fig11}
\end{figure*}

\begin{figure}[tb]
%\centering
\includegraphics[width=0.5\columnwidth]{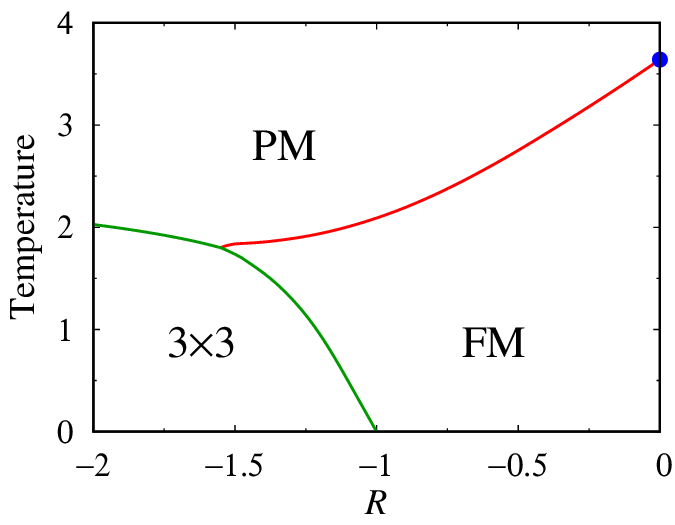}
\caption{Finite temperature phase diagram for $J_1 = 1$ and $R = J_2 / J_1$ in GTL2. PM and FM represent the paramagnetic and ferromagnetic phases, respectively. The exact value~\cite{Wannier50,Codello10} of $T_c=4/\log(3)$ at $R=0$ is indicated by the solid circle. The error bar is comparable to the line thickness.}
\label{fig12}
\end{figure}

\subsection{$J_1$-$J_2$ Ising model in GTL2}

Figure~\ref{fig8} displays the entropy profile as a function of $E_1$ and $E_2$ for the $J_1$-$J_2$ Ising model in GTL2. The four corner points $(E_1=3, E_2=1)$, $(E_1=5/3, E_2=-1/3)$, $(E_1=-1, E_2=1)$, and $(E_1=-1, E_2=-1/3)$ represent the ferromagnetic, ferrimagnetic 3$\times$3, and two kinds of partial spin liquid ground states (PSL1 and PSL2), respectively. In the partial spin liquid (PSL) ground state, a portion of spins remains fluctuating even at zero temperature while the other spins are frozen~\cite{Yu16}. The parameter regimes of $J_1$ and $J_2$ and typical ground state spin configurations for the 3$\times$3 and PSL ground states are illustrated in figure~\ref{fig9}. It is worth noting that next-nearest-neighbor interaction exists only between spins with six next-nearest-neighbors, forming the triangular lattice with lattice constant $\sqrt{3}$ in GTL2. In the 3$\times$3 phase, spins in one of tripartite sublattices of next-nearest-neighbor interaction lattice align up ({\it resp.} down) whereas all the other spins align down ({\it resp.} up). Thus, the magnetization is $7/9$ and the residual entropy is $\log(6)$ in the 3$\times$3 ground state.

The PSL1 ground state is similar to that of the conventional triangular Ising antiferromagnet. All the elemental triangles must adhere to the 2-1 rule, which means that one or two spins among the three spins of every triangle should point up and the other two or one spin should point down~\cite{Yu15PRE}. However, an additional restriction exists: all the six-next-nearest-neighbor spins have the same spin direction. Thus, the residual entropy should be smaller than the antiferromagnetic Ising model in the triangular lattice. Although it is difficult to calculate the residual entropy of the PSL1 phase exactly, it can be estimated using the Pauling's approximation~\cite{Pauling35}. Consider a hexagon composed of one six-nearest-neighbor spin in the center and six no-nearest-neighbor spins at the vertices. The direction of the central spin is fixed. Among $2^6$ spin configurations for the six spins at the vertices, only 18 configurations satisfy the 2-1 rule. Since the whole lattice can be viewed as the tiling of edge-sharing hexagons, the residual entropy of the PSL1 phase is estimated as
\begin{eqnarray}
    S_0 = \log \left[ 2^N \left( \frac{18}{2^6} \right)^{N/3} \right] 
    = \frac{2}{3} \log\left[\frac{3}{2}\right] N \approx (0.270310) N .
\end{eqnarray}
Here the exponent $N/3$ indicates that there are $N/3$ hexagons in the lattice with $N$ spins. This estimate is consistent with our results obtained by the Wang--Landau method ($S_0 \approx (0.291214) N$; see figure~\ref{fig10}) within 7\%. Regarding the PSL2 ground state, the directions of the no-next-nearest-neighbor spins are fixed and the six-next-nearest-neighbor spins form a triangular lattice with their next-nearest-neighbors. Consequently, the residual entropy of the PSL2 phase is $S_0 = (\sigma_0^{T}/3) N$, where $\sigma_0^{T}\approx 0.323066$ is the residual entropy per spin in the triangular Ising antiferromagnet~\cite{Wannier50}. As shown in figure~\ref{fig10}, this formula explains our results very well.

Figure~\ref{fig11} shows magnetization ($m$ and $m_{3\times 3}$), magnetic susceptibility ($\chi$ and $\chi_{3\times 3}$), Binder cumulant ($U$ and $U_{3\times 3}$), and specific heat ($C$) as a function of temperature ($T$) for $J_1=1$ and $J_2=R$ with $R = -0.5$, $-1.1$, and $-1.8$. The order parameter of the 3$\times$3 phase is defined by 
\begin{eqnarray}
m_{3\times 3} = \frac{1}{N} &\left[ \mathrm{Max}\left\{ \left|-M_0^{(6)} + M_1^{(6)} + M_2^{(6)} + M^{(0)} \right|, \right. \right. \nonumber \\
&~~~~~~~~~~\left| M_0^{(6)} - M_1^{(6)} + M_2^{(6)} + M^{(0)} \right|, \nonumber \\
&~~~~~~~~~\,\left. \left.  \left| M_0^{(6)} + M_1^{(6)} - M_2^{(6)} + M^{(0)} \right|  \right\} \right]
\end{eqnarray}
where $M_i^{(6)}$ represents the total magnetization of $i$th tripartite sublattice of the next-nearest-neighbor lattice composed of the six-next-nearest-neighbor spins and $M^{(0)}$ is the total magnetization of the no-next-nearest-neighbor spins. The order parameter $m_{3\times 3}$ becomes 1 in the 3$\times$3 ground state and $m_{3\times 3} = 7/9$ for the ferromagnetic phase. Corresponding susceptibility ($\chi_{3\times 3}$) and Binder cumulant ($U_{3\times 3}$) are defined by replacing $m$ with $m_{3\times 3}$ in (\ref{eq_sus})--(\ref{eq_U}). For $R=-0.5$ and $R=-1.8$, typical continuous phase transitions occur from the paramagnetic phase into the ferromagnetic and 3$\times$3 phases, respectively. 
Interestingly, as evidenced by the double-peak structures in specific heat and $\chi_{3\times 3}$ shown in Fig.~\ref{fig11}, successive phase transitions occur in the case of $R=-1.1$. As the temperature decreases, the magnetization ($m$) increases up to $0.86$ at the higher transition temperature $T_c$ and then decreases to $7/9$ at the lower $T_c$, while $m_{3\times 3}$ increases in two steps: from zero to $7/9$ and from $7/9$ to $1$. The Binder cumulant $U_{3\times 3}$ for the 3$\times$3 phase increases monotonically from zero to $2/3$ at the higher $T_c$, similar to the continuous ferromagnetic transition of the unfrustrated Ising model. In the disordered phase, the Binder cumulant is zero, whereas it becomes $2/3$ in ordered phases where the corresponding order parameter is steady with small variance. At the lower $T_c$, $U_{3\times 3}$ slightly drops before increasing again to $2/3$, indicating a phase transition between ordered phases. Thus, we conclude that the two phase transitions are from the paramagnetic phase into the ferromagnetic phase at the higher $T_c$ and from the ferromagnetic phase into the 3$\times$3 phase at the lower $T_c$.
The former is associated with the $Z_2$ symmetry breaking, and the latter with the translational symmetry breaking.
We examined the energy histogram near the two phase transitions, but no double-peak structure was found.

The finite-temperature phase diagram is presented in figure~\ref{fig12}. The critical temperature was determined by the crossing of the Binder cumulant ($U$ and $U_{3\times 3}$) data for the two largest lattices studied in this work ($L=36$ and $L=42$). The ferromagnetic critical temperature decreases as $R$ decreases, and double-transition behavior is observed in $-1.55 < R < -1$. For $R < -1.55$, the ferromagnetic phase disappears and there exists only one phase transition into the 3$\times$3 phase.

\section{Summary \label{Sec_Summary}}

We studied the frustrated $J_1$-$J_2$ Ising model in two kinds of generalized triangular lattices employing the Wang--Landau method combined with finite-size scaling techniques. Zero-temperature phase diagrams were constructed across the entire parameter space of $J_1$ and $J_2$ from the entropy profile. In GTL1, which is anisotropic, three distinct super-antiferromagnetic ground states with stripe structures were identified. On the other hand, the 3$\times$3 and two kinds of partial spin liquid ground states were observed in GTL2, which has non-regular next-nearest-neighbor interaction. In the partial spin liquid ground states, residual entropy was verified to be proportional to the number of spins.

Finite-temperature phase diagrams were also established as a function of $R=J_2/J_1$ for $J_1=1$ and $J_2<0$. In GTL1, the critical temperature $T_c$ of continuous phase transition from the paramagnetic phase into the ferromagnetic phase decreases monotonically as $R$ decreases, ranging from $T_c=4/\log(3)$ at $R=0$ to $T_c=0$ at $R=-1$. For $R<-1$, the first-order phase transition into the stripe phase appears. In GTL2, $T_c$ into the ferromagnetic ground state decreases as $R$ decreases until it meets the phase boundary of the 3$\times$3 state at $R=-1.55$. Within the range $-1.55<R<-1$, two successive transitions occur: a transition from the paramagnetic phase to the ferromagnetic phase at the higher critical temperature followed by another transition from the ferromagnetic phase to the 3$\times$3 phase at the lower critical temperature.

\ack
This work was supported by the National Research Foundation of Korea(NRF) grant funded by the Korea government(MSIT) (No. 2021R1F1A1052117) and by the GIST Research Project grant funded by the GIST in 2024.

\section*{References}
\bibliography{main}

\end{document}